\documentclass[
  twocolumn,
  superscriptaddress,
  amsmath,amssymb,
  aps,
  nofootinbib,
  reprint
]{revtex4-2}

\usepackage{orcidlink}    % orcid
\usepackage{graphicx}     % figures
\usepackage{dcolumn}      % decimal-aligned columns
\usepackage{bm}           % bold math
\usepackage{mathtools}    % math extras (paired delimiters, etc.)
\usepackage{slashed}      % \slashed{p}
\usepackage{booktabs}     % nice tables
\usepackage{array}        % column types
\usepackage{acro}         % acronyms
\usepackage{hyperref}
\usepackage[nameinlink]{cleveref}
\usepackage{xcolor}
\usepackage{etoolbox}
\usepackage[normalem]{ulem} % for \sout
\definecolor{Red}{rgb}{1,0,0}
\definecolor{Green}{rgb}{0,0.6,0}
\definecolor{Blue}{rgb}{0,0,1}
\definecolor{Cyan}{rgb}{0,1,1}
\definecolor{Magenta}{rgb}{1,0,1}
\definecolor{Yellow}{rgb}{1,1,0}
\definecolor{Orange}{rgb}{1,0.5,0}
\definecolor{Purple}{rgb}{0.5,0,0.5}
\definecolor{Teal}{rgb}{0,0.5,0.5}
\definecolor{Gray}{gray}{0.5}
\definecolor{LightGray}{gray}{0.93}
\definecolor{Black}{gray}{0}
\definecolor{White}{gray}{1}

\colorlet{RedLight}{Red!25!White}
\colorlet{RedDark}{Red!70!Black}
\colorlet{GreenLight}{Green!25!White}
\colorlet{GreenDark}{Green!70!Black}
\colorlet{BlueLight}{Blue!25!White}
\colorlet{BlueDark}{Blue!70!Black}
\colorlet{CyanLight}{Cyan!25!White}
\colorlet{CyanDark}{Cyan!70!Black}
\colorlet{MagentaLight}{Magenta!25!White}
\colorlet{MagentaDark}{Magenta!70!Black}
\colorlet{YellowLight}{Yellow!35!White}
\colorlet{YellowDark}{Yellow!70!Black}
\colorlet{OrangeLight}{Orange!25!White}
\colorlet{OrangeDark}{Orange!70!Black}
\colorlet{PurpleLight}{Purple!25!White}
\colorlet{PurpleDark}{Purple!70!Black}
\colorlet{TealLight}{Teal!25!White}
\colorlet{TealDark}{Teal!70!Black}
\colorlet{GrayLight}{Gray!25!White}
\colorlet{GrayDark}{Gray!70!Black}
\newcolumntype{C}{>{$}c<{$}}
\AtBeginDocument{
  \heavyrulewidth=.16em
  \lightrulewidth=.1em
  \cmidrulewidth=.03em
  \belowrulesep=.4ex
  \belowbottomsep=0pt
  \aboverulesep=.4ex
  \abovetopsep=0pt
  \cmidrulesep=\doublerulesep
  \cmidrulekern=.5em
  \defaultaddspace=.5em
}
\crefname{section}{Sec.\!}{Secs.\!}
\crefname{figure}{Fig.\!}{Figs.\!}
\crefname{equation}{}{}
\crefname{table}{Tab.\!}{Tabs.\!}
\crefname{appendix}{App.\!}{Apps.\!}

\hypersetup{
    colorlinks=true,       % false: boxed links; true: colored links
    linkcolor=blue,        % color of internal links (equations, figures, etc.)
    citecolor=blue,        % color of links to bibliography
    filecolor=magenta,     % color of file links
    urlcolor=cyan          % color of external links
}
\newcommand{\dd}{\mathrm{d}}

\DeclareAcronym{QFT}{short=QFT,long=quantum field theory,short-plural-form=QFTs,long-plural-form=quantum field theories}
\DeclareAcronym{EFT}{short=EFT,long=effective field theory,short-plural-form=EFTs,long-plural-form=effective field theories}
\DeclareAcronym{UV}{short=UV,long=ultraviolet}
\DeclareAcronym{IR}{short=IR,long=infrared}
\DeclareAcronym{SM}{short=SM,long=Standard Model}
\DeclareAcronym{RG}{short=RG,long=renormalization group}
\DeclareAcronym{FRG}{short=FRG,long=functional renormalization group}
\DeclareAcronym{AS}{short=AS,long=asymptotic safety}
\DeclareAcronym{QCD}{short=QCD,long=quantum chromodynamics}
\DeclareAcronym{QED}{short=QED,long=quantum electrodynamics}

\DeclareAcronym{MS}{short=MS,long=minimal subtraction}
\DeclareAcronym{MOM}{short=MOM,long=momentum subtraction}
\DeclareAcronym{DR}{short=DR,long=dimensional regularization}

\DeclareAcronym{GUT}{short=GUT,long=grand unified theory, short-plural-form=GUTs, long-plural-form=grand unified theories}
  {\list{}{\leftmargin=0.0in\rightmargin=0.0in}\item[]}
  {\endlist}
  
\makeatletter
\def\bibsection{%
  \par
  \begingroup
  \baselineskip26\p@
  \bib@device{\columnwidth}{200\p@}%
  \endgroup
  \nobreak\@nobreaktrue
  \addvspace{19\p@}%
}
\makeatother
\begin{document}

\title{Smooth Threshold Effects from Dimensional Regularization}

\author{Yannick Kluth\,\orcidlink{0000-0002-8126-7668}}
\email{yannick.kluth@utoronto.ca}
\affiliation{
 Department of Physics, University of Toronto, Toronto, ON M5S 1A7, Canada
}

\date{\today}

\begin{abstract}
We suggest a non-minimal renormalization scheme based on dimensional regularization that naturally incorporates threshold effects of heavy particles. By renormalizing couplings and masses to subtract all poles in $d \geq 4$, the resulting scheme is mass-dependent and circumvents shortcomings of mass-independent schemes like minimal subtraction. At the same time, many advantages of minimal subtraction such as gauge independence are retained. Through explicit one-loop computations in QCD, we demonstrate that this scheme reduces to minimal subtraction at high energies while providing smooth transitions at particle thresholds and implementing the Appelquist-Carazzone theorem. Potential future applications and extensions are discussed.
\end{abstract}

\maketitle

\section{Introduction}
The \ac{SM} of particle physics contains various mass scales. While particles are effectively massless above these scales, the Appelquist-Carazzone theorem  ensures that heavy degrees of freedom decouple when the energy scale falls below their threshold \cite{Appelquist:1974tg}. In this case, heavy particles can be integrated out leading to a low energy \ac{EFT} containing only light degrees of freedom. The transition between these regimes is described by threshold effects; heavy degrees of freedom \textit{smoothly} decouple and lead to a transition between the high and low energy theories.

Although the decoupling of heavy particles is conceptually well-understood, its implementation in perturbative \ac{RG} functions proves non-trivial. While \ac{MS} with \ac{DR} allows for high precision computations due to its technical simplicity \cite{vanRitbergen:1997va,Czakon:2004bu,Baikov:2016tgj,Schnetz:2016fhy,Herzog:2017ohr}, 
it is oblivious to mass scales. As such, threshold effects are ignored and running couplings fail to manifest the decoupling of heavy degrees of freedom. While this is negligible at high energies, large logarithms are generated  as threshold effects become relevant. These require resummation to prevent a breakdown of perturbation theory \cite{Bernreuther:1994nw,Beneke:2011mq}. Although the decoupling theorem can be implemented manually into \ac{MS} by matching to a low energy \ac{EFT} \cite{Burgess:2007pt}, this leads to discontinuities in \ac{RG} functions at particle thresholds \cite{Bernreuther:1981sg,Chetyrkin:1997un,Chetyrkin:2000yt,Chetyrkin:2005ia,Schroder:2005hy}.  
Hence, a scheme that naturally implements the decoupling of heavy particles would be much preferred \cite{Brodsky:1998mf}.

To define \ac{RG} functions that manifest the decoupling theorem, it is necessary to go beyond \ac{MS} and employ non-minimal subtractions. One example is \ac{MOM} \cite{Georgi:1976ve}, in which couplings are renormalized by projecting onto vertices at given momentum configurations. However, \ac{MOM} is computationally more demanding since it relies on finite parts. Moreover, ambiguities arise such as the freedom of choosing different vertices and projectors to define running couplings. Thus, different versions of \ac{MOM} can be defined, leading to different results for \ac{RG} functions \cite{Celmaster:1979km,Chetyrkin:2000fd,Gracey:2011vw,Gracey:2011pf,Bednyakov:2020cdf}. While many of these ambiguities can be avoided by using the background field method \cite{Abbott:1980hw,Rebhan:1985yf,Chetyrkin:2008jk}, choices for momentum configurations remain non-trivial and results retain a dependence on gauge parameters \cite{Jegerlehner:1998zg,Zeng:2020lwi}.

An alternative approach to constructing non-minimal renormalization schemes is provided by \ac{DR} via the subtraction of poles in other dimensions than $d = 4$. Poles in $d < 4$ can be interpreted as power-law divergences \cite{Veltman:1980mj} and their inclusion in the renormalization process is crucial for \ac{UV} phenomena in theories where these arise \cite{Castellani:1982dk,Jack:1990pz,Al-Sarhi:1990nmv,Al-sarhi:1991gdi,Kaplan:1998tg,Kluth:2024lar,cckl}. However, \ac{DR} also generates poles in $d > 4$, which could be included in the renormalization process \cite{Weinberg:1980gg}. Such a scheme was recently applied in \cite{Falls:2024noj}, where poles in $d \leq 4$ as well as $d > 4$
were considered to study the \ac{UV} behaviour of quantum gravity. In this work, we take a slightly different conceptual angle and construct a renormalization scheme by subtracting all poles in $d \geq 4$ and show that this is particularly useful for \ac{IR} phenomena. We demonstrate that the resulting scheme is mass-dependent and naturally describes smooth threshold effects of massive particles. Moreover, at large energies, the subtraction of poles in $d > 4$ becomes irrelevant and we recover \ac{MS}. Since the non-minimal subtractions in our scheme only involve poles in $d > 4$, they carry over many advantages of \ac{MS}. In particular, they are free of ambiguities that arise in \ac{MOM}, and manifestly gauge independent.

This letter is organized as follows. First, we motivate the inclusion of \ac{UV} poles in higher dimensions before outlining the implementation of this renormalization procedure. The framework is applied to \ac{QCD} with massive fermions at one-loop. \ac{RG} functions for couplings and masses are computed and their sensitivity to thresholds is established. We conclude by discussing potential future applications.

\section{IR Physics from Higher Dimensions}
\label{sec:scheme}
The notion that \ac{IR} physics can be inferred from \ac{UV} structures in higher dimensions was first recognized in the context of the Wilson-Fisher fixed point \cite{Wilson:1971dc}. This non-trivial \ac{IR} fixed point is a feature of the scalar $\phi^4$ theory in $d=3$, which is super-renormalizable and asymptotically free, but shows non-trivial \ac{IR} behaviour. Since the $\phi^4$ coupling has positive mass dimension in $d=3$, it does not receive logarithmic divergences or poles. Therefore, the \ac{MS} $\beta$-function is governed solely by its classical scaling dimension, and fails to capture \ac{IR} effects of the Wilson-Fisher fixed point. This limitation is very similar to the impact of heavy particle thresholds, where \ac{MS} does not naturally incorporate the decoupling of massive degrees of freedom. Both scenarios showcase the limitations of \ac{MS} to non-trivial effects in the \ac{IR}.

In the context of the Wilson-Fisher fixed point, one way of circumventing such shortcomings is the $\varepsilon$-expansion.
Instead of computing the running coupling directly in $d = 3$ with \ac{MS}, the theory is renormalized in $d = 4 - \varepsilon$, and then analytically continued to $\varepsilon \to 1$. In this way, the \ac{UV} pole structure in $d = 4$ is leveraged to infer long-distance effects of the theory in $d = 3$. Even though this approach lacks a rigorous proof of convergence for $\varepsilon \to 1$, it turned out remarkably effective. Results for critical exponents are in exceptional agreement with complementary techniques such as lattice computations or the conformal bootstrap \cite{Guida:1998bx,Hasenbusch:2010hkh,El-Showk:2014dwa,Kos:2016ysd,Kompaniets:2017yct,Chang:2024whx}. 

The success of the $\varepsilon$-expansion suggests that poles of \ac{DR} in higher dimensions could be useful in more general terms. A generalization of analytically continuing from e.g. four dimensional spacetime to three dimensions can be provided by using a renormalization scheme that includes non-minimal subtractions. In this work, we will consider the following renormalization scheme for a theory in $d_*$ dimensional spacetime:
\begin{quote}
    \centering
    \itshape
    Renormalize couplings and masses by minimally subtracting all UV divergences in $d \geq 4$.
\end{quote}
Thus, in addition to the poles in $d=d_*$, which are the only terms included in \ac{MS}, we also include non-minimal subtractions that are finite in $d \to d_*$. As we will see, these additional contributions will be useful to capture relevant \ac{IR} effects such as thresholds of massive particles.

Note that our scheme can be applied to both power-counting non-renormalizable as well as renormalizable theories such as the \ac{SM}. In the latter case, we restrict the subtraction of poles in higher dimensions to those that are proportional to the operators present in the underlying action. Contributions to higher dimensional operators that are not part of the action (e.g. SMEFT operators like $(\phi^\dagger \phi)^3$) are ignored.

The dimensions $d$ in which relevant divergences occur that we will include can be determined a priori by dimensional analysis. In a massless theory, any given operator $\mathcal{O}_i$ generates a pole only in the critical dimension $d_{i, \text{crit}}$ where it is marginal (e.g., $d_\text{crit}=4$ for $\phi^4$). Without a mass parameter, all poles in $d > d_{i, \text{crit}}$ must be proportional to operators that are not included in the Lagrangian and can be ignored according to our scheme.
However, the introduction of a mass fundamentally alters this behavior. Given any mass scale $M$, one can construct monomials of the form $M^N \mathcal{O}_i$, which are marginal in $d = d_{i, \text{crit}} + N \geq d_{i, \text{crit}}$. For instance, while the $\varepsilon$-expansion leverages the pole of $\phi^4$ in its critical dimension at $d=4$, a massive theory would also generate poles proportional to $M^2 \phi^4$ in $d=6$, $M^4 \phi^4$ in $d=8$, and so on. Consequently, in the presence of masses, our scheme accounts for an infinite tower of higher-dimensional poles that all contribute to the renormalization of couplings and masses. As we will see by an explicit calculation below, it will turn out that these contributions converge and can be summed, giving rise to well defined results for renormalization constants.

\section{Threshold Effects in QCD}

To demonstrate its effectiveness, we apply the scheme of \cref{sec:scheme} to \ac{QCD}. We consider a $SU(N_c)$ gauge theory including $n_F$ massive fermions $\Psi^r$ with dimensionful bare masses $M_{0, r}$,
\begin{equation}
    \mathcal{L} = - \frac{1}{4 g_0^2} F^{a \mu \nu} F^a_{\mu \nu} + \sum_{r = 1}^{n_F} \left[ \bar{\Psi}^r i \slashed{D} \Psi^r - M_{0, r} \bar{\Psi}^r \Psi^r \right]\, ,
    \label{eqn:qcdLagrangian}
\end{equation}
where $g_0$ denotes the bare gauge coupling, and the index $r$ is used to distinguish different quark flavours.
To compute the necessary poles in \ac{DR} we will utilize the background field method and add the usual background field gauge fixing and ghost terms, see \cite{Abbott:1980hw} for details. However, just as in \ac{MS}, all results can be equivalently computed from standard Feynman diagrams.\footnote{This equivalence between the background field method and standard Feyman diagrams is usually not realized for schemes that rely on finite parts such as \ac{MOM}.}

Since the \ac{UV} structure of \ac{QCD} in $d = 4$ is well understood, we will only focus on new contributions arising from poles in $d>4$. As explained above, such poles must be associated with the fermion masses, since none of the operators included in \cref{eqn:qcdLagrangian} is marginal in $d > 4$. As such, we are only required to compute fermionic contributions in higher dimensions. Any contributions that are fermion independent are also mass independent, and, therefore, will not contribute to the renormalization of $g_0$ or $M_{0,r}$.

\subsection{Coupling Renormalization}
\begin{figure}
    \centering
    \includegraphics[width=0.7\linewidth]{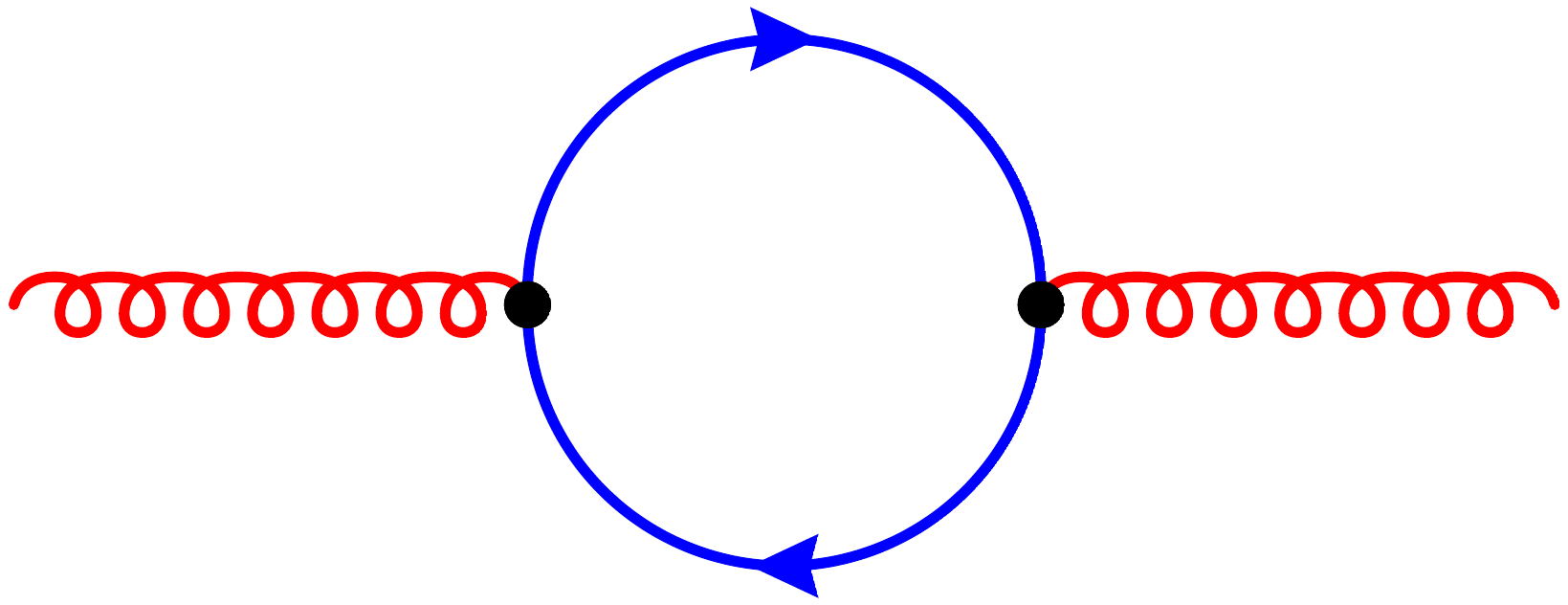}
    \caption{Fermionic contribution to the background gluon two-point function at one-loop. All Feynman diagrams in this publication were drawn with \texttt{FeynGame} \cite{Harlander:2020cyh,Harlander:2024qbn,Bundgen:2025utt}.}
    \label{fig:AA}
\end{figure}

Due to background gauge invariance, the renormalization of $g_0$ can be determined by computing the background gluon two-point function. At one-loop, the single relevant fermionic Feynman diagram is shown in \cref{fig:AA}, where we implicitly assume a summation of all fermion flavours in the fermion loop. Note that our purpose is to only compute the renormalization of $g_0$ and not of any other higher dimensional operators induced in higher dimensions. For this reason, we only compute \cref{fig:AA} up to second order in a momentum expansion. Contributions that are of higher order in momenta correspond to higher dimensional operators not included in \cref{eqn:qcdLagrangian}, and, thus, will be ignored.

Since \cref{fig:AA} is free of \ac{IR} divergences due to the fermion masses, we can perform a naive expansion in external momenta $p_\mu$ to identify \ac{UV} poles. Expanding to second order in $p_\mu$, each individual fermion flavour gives a contribution from \cref{fig:AA} of the form,
\begin{equation}
    \begin{split}
        \Pi_{\mu \nu}^{f, a b} =&\, \frac{i \pi}{3} \mathcal{T}^{a b}_{\mu \nu} \sum_{r = 1}^{n_F}
        \frac{
            (d-2) M_{0, r}^{d-4}
        }{
            (4 \pi)^{d/2} \Gamma \left(\frac{d}{2}\right) \sin \left(\frac{\pi d}{2}\right)
        } + \mathcal{O} (p^4) \, .
    \end{split}
    \label{eqn:AAdiv}
\end{equation}
with $\mathcal{T}^{a b}_{\mu \nu} = \delta^{a b} (\eta_{\mu \nu} p^2 - p_\mu p_\nu)$. Note that \cref{eqn:AAdiv} is gauge parameter independent.
\ac{UV} poles of \cref{eqn:AAdiv} occur in $d = 4 + 2n$ with $n \in \mathbb{N}_0$.\footnote{While the factor of $\sin (\tfrac{\pi d}{2})$ in the denominator of \cref{eqn:AAdiv} leads to poles in any even $d$ (including $d<4$) all poles in $d < 4$ are compensated by factors of $d - 2$ in the numerator or $\Gamma (\tfrac{d}{2})$ in the denominator.} The coupling $g_0$ is renormalized according to
\begin{equation}
    g_0 = \mu^{\tfrac{4 - d}{2}} Z_g g \, ,
\end{equation}
with $\mu$ the sliding scale, $g$ the renormalized coupling, and the renormalization constant $Z_g$ is given by
\begin{equation}
    Z_g = 1 + g^2 \left[  \delta Z_g^\text{YM} + \delta Z_g^f \right] + \mathcal{O} (g^4) \, ,
\end{equation}
where we have separated the one-loop fermionic contributions $\delta Z_g^f$ from the pure Yang-Mills contributions at one-loop ($\delta Z_g^\text{YM}$).
We determine $\delta Z_g^f$ to subtract all poles of \cref{eqn:AAdiv}, giving rise to
\begin{equation}
    \delta Z_g^\text{f} = \frac{2}{3 (4 \pi)^2} \sum_{r = 1}^{n_F} \sum_{n = 0}^\infty \frac{\big(-m_r^2\big)^n}{n! (2n + 4 - d)} \, ,
    \label{eqn:ZgFermion2}
\end{equation}
with the dimensionless renormalized mass $m_r$ defined by
\begin{equation}
    \mu \,Z_m m_r = \frac{M_{0, r}}{\sqrt{4 \pi}} \, .
    \label{eqn:renMass}
\end{equation}
Although subtracting all poles in higher dimensions generates an infinite series in $m_r$, \cref{eqn:ZgFermion2} converges and can be expressed in terms of the lower incomplete $\Gamma$-function,
\begin{equation}
    \delta Z_g^\text{f} = \frac{1}{3 (4 \pi)^2} \sum_{r = 1}^{n_F} m_r^{d-4} \gamma
    \left(\tfrac{4 - d}{2},m_r^2\right) \, ,
    \label{eqn:ZgFermion}
\end{equation}
where
\begin{equation}
    \gamma(s, x) = \int_0^x \dd t \, t^{s - 1} e^{-t} \, .
\end{equation}

The fermionic contributions \cref{eqn:ZgFermion} are then supplemented by the usual \ac{UV} divergences in $d = 4$ of pure Yang-Mills theories encoded in $\delta Z_g^\text{YM}$. Combining both, and using $\alpha = g^2/(4 \pi)$, we find for the $d$-dimensional $\beta$-function,
\begin{equation}
    \begin{split}
        \beta_\alpha =&\, (d - 4) \, \alpha + \frac{\alpha^2}{4 \pi} \left[ - \frac{22}{3} C_A + \frac{4}{3} \sum_{r = 1}^{n_F} e^{- m_r^2}  \right] \,,
    \end{split}
    \label{eqn:betares}
\end{equation}
where $\beta_\alpha \equiv \mu \tfrac{\partial \alpha}{\partial \mu}$, and $C_A = N_c$ is the Casimir operator of $SU(N_c)$ in the adjoint representation. We identify the contribution of each quark flavour as
\begin{equation}
    \beta_r \equiv \frac{\alpha^2}{4 \pi} \frac{4}{3} e^{-m_r^2} \, .
    \label{eqn:quarkOneLoop}
\end{equation}
Thus, the inclusion of \ac{UV} divergences from higher dimensions leads to exponential factors controlled by the dimensionless masses $m_r$. 
Assuming the classical scaling of $m_r$ to be dominant over quantum corrections,  we infer from \cref{eqn:renMass} that $m_r \to 0$ for $\mu \gg M_r$. Thus, at large energies the exponential factors become one and \cref{eqn:betares} approaches the well-known \ac{MS} results at one-loop. In the \ac{IR} the dimensionless masses $m_r$ diverge like $1/\mu$. As such, fermionic contributions become exponentially suppressed, which establishes a smooth decoupling of heavy masses.

\begin{figure}
    \centering
    \includegraphics[width=0.9\linewidth]{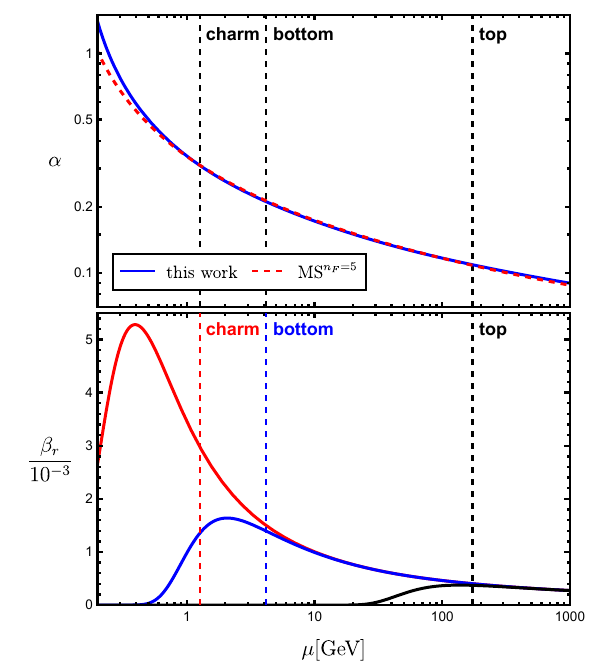}
    \caption{The top panel shows the \ac{QCD} coupling $\alpha$ obtained from \cref{eqn:betares} in comparison to the \ac{MS} running at one-loop with $n_F = 5$ and initial conditions set at $M_Z =  91.1876 \, \text{GeV}$ with $\alpha(M_Z) = 0.1184$.
    The bottom panel illustrates contributions of different quarks to the \ac{QCD} $\beta$-function following \cref{eqn:quarkOneLoop}. Vertical lines indicate quark masses; contributions peak below those scales due to the \ac{IR} growth of the \ac{QCD} coupling.}
    \label{fig:qcdRunning}
\end{figure}

As a concrete example, we apply the running described by \cref{eqn:betares} to the case of full \ac{QCD} including all quarks in \cref{fig:qcdRunning}. For simplicity, we assume quark masses to follow classical running. This is a good approximation since the classical scaling of masses is much larger than quantum induced anomalous dimensions, which we will also verify below. 

At energies far above any mass thresholds, all fermion flavours contribute equally to the running, on par with results from \ac{MS}. As the energy scale is lowered, each fermion flavour decouples when the energy falls below their mass and stops contributing to the running. Within the perturbative regime, this leads to smooth thresholds for the top, bottom, and charm quarks, before $\alpha$ becomes strongly coupled in the \ac{IR}. Defining the \ac{QCD} scale parameter $\Lambda_\text{QCD}$ by the \ac{IR} pole of the one-loop running obtained from \cref{eqn:betares}, we find
\begin{equation}
    \Lambda_\text{QCD}^\text{1-loop} = 121.7 \, \text{MeV} \, .
\end{equation}
Using \ac{MS} at one-loop with $n_F = 5$ and without imposing matching conditions to implement decouplings of heavy quarks, we would find $\Lambda_\text{QCD}^\text{MS} = 89.9 \, \text{MeV}$. This smaller value for $\Lambda_\text{QCD}$ obtained from \ac{MS} is explained by the fact that fermionic contributions decrease the running coupling towards the \ac{IR}. In \ac{MS}, all quarks contribute equally at all energy scales, leading to a larger effective number of quarks being present at low energies compared to more realistic schemes like \cref{eqn:betares}, where the decoupling theorem is realized.

\subsection{Mass Renormalization}
\begin{figure}
    \centering
    \includegraphics[width=0.7\linewidth]{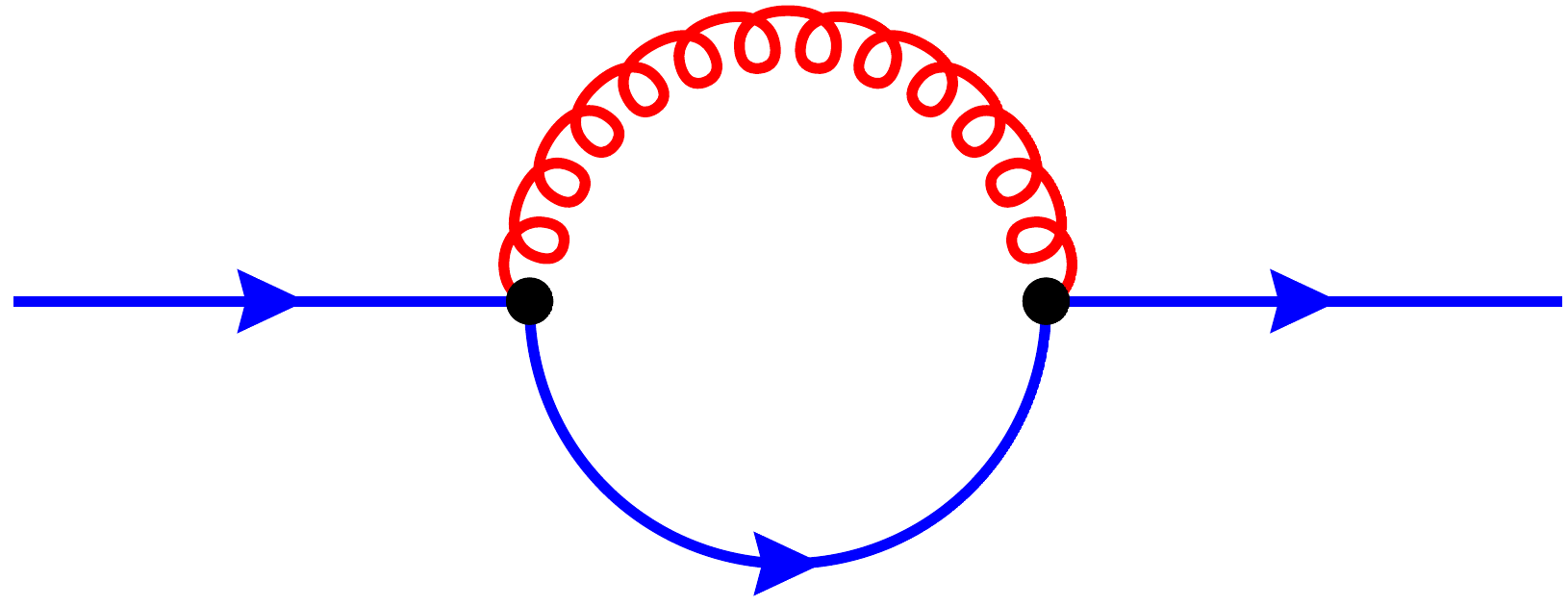}
    \caption{Fermionic two-point function at one-loop.}
    \label{fig:psipsibar}
\end{figure}

The scheme in \cref{sec:scheme} can also be applied to the renormalization of quark masses, by computing the fermionic two-point function.
At one-loop, this results in a single Feynman diagram shown in \cref{fig:psipsibar}. One subtlety arising in this computation is the non-trivial wave-function renormalization for the quarks, which can become momentum dependent in $d > 4$. To compute the correct renormalization for quark masses, it is necessary to distinguish contributions to the mass from contributions to the wave-function renormalization. This can be done by computing the wave-function renormalization explicitly. However, to avoid computing momentum-dependent wave-function renormalizations in $d > 4$, we can instead project on-shell to eliminate any off-shell contributions. In the case of the fermionic two-point function, this is done by fixing external momenta to be on-shell, i.e. $p^2 = M_{0, r}^2$. With that, the on-shell self energy from \cref{fig:psipsibar} becomes
\begin{equation}
    \begin{split}
        \Sigma_{i j} \Big|_\text{on-shell} =&\, -i \pi C_F \delta_{i j} 
        \frac{(d-1) g_0^2 M_{0, r}^{d-3} \csc \left(\frac{\pi  d}{2}\right)}{(4 \pi)^{d/2} (d-3) \Gamma \left(\frac{d}{2}-1\right)} \, ,
    \end{split}
    \label{eqn:psibarPsiRes}
\end{equation}
with $C_F = (N_c^2 - 1)/(2 N_c)$ being the Casimir operator in the fundamental representation, and we note that \cref{eqn:psibarPsiRes} is again gauge-parameter independent.\footnote{To obtain this gauge independence, it is crucial to distinguish contributions to wave-function renormalizations from mass renormalizations.}
Similarly to \cref{eqn:AAdiv}, this contains poles at $d = 2n + 4$, with $n \in \mathbb{N}_0$, while also being divergent in $d = 3$. Crucially, the pole in $d = 3$ of \cref{eqn:psibarPsiRes} does \textit{not} signify a \ac{UV} divergence, but stems entirely from an \ac{IR} divergence. This occurs because \cref{fig:psipsibar} develops \ac{IR} divergences for on-shell momenta in $d \leq 3$, while being \ac{IR} finite in $d \geq 4$. Indeed, explicitly regularizing all \ac{IR} divergences—for instance, via \ac{IR} rearrangement \cite{Vladimirov:1979zm,Misiak:1994zw,Chetyrkin:1997fm}—eliminates the $d = 3$ pole in \cref{eqn:psibarPsiRes} while leaving the \ac{UV} poles at $d \geq 4$ unaffected. Consequently, the $d = 3$ pole should be excluded from the renormalization process. Moreover, while power-counting suggests potential linear \ac{UV} divergences to be present in the quark masses which would correspond to poles at $d = 3$, chiral symmetry protects the mass from such contributions. Thus, \ac{UV} poles must be absent in $d = 3$.

To subtract divergences of \cref{eqn:psibarPsiRes} in $d \geq 4$, we determine the mass renormalization from \cref{eqn:renMass} as
\begin{equation}
    \begin{split}
        Z_{m_r} =&\, \frac{- 2 g^2 C_F}{(4 \pi)^2} \sum_{n = 0}^\infty \frac{2n+3}{(2n+1) n!} \frac{(-m_r^2 )^n}{2n + 4 - d} \\
        =&\, \frac{g^2 C_F}{(4 \pi)^2} \frac{F(m_r^2,3) - F(m_r^2,d)}{d - 3} \, , \\
    \end{split}
    \label{eqn:ZmSum}
\end{equation}
with
\begin{equation}
    F(m_r^2, d) = (d - 1) m_r^{d - 4} \, \gamma \left( \tfrac{4 - d}{2}, m_r^2 \right) \, .
\end{equation}
Note that the pole at $d = 3$ in \cref{eqn:ZmSum} is spurious, i.e. $Z_{m_r}$ is finite in $ d= 3$.
With the renormalization constant at hand, we determine the one-loop mass running for any fermion in this theory to be of the form
\begin{equation}
    \frac{\beta_{m_r}}{m_r} = - 1 - \frac{3 \alpha}{2 \pi} C_F \, \zeta (m_r) \, ,
    \label{eqn:betaMass}
\end{equation}
with $\beta_{m_r} \equiv \mu \tfrac{\partial m_r}{\partial \mu}$, and the threshold profile given by
\begin{equation}
    \zeta (m_r) = \frac{1}{3} \left[ e^{-m_r^2} + \frac{\gamma \left(\tfrac{1}{2},m_r^2\right)}{m_r} \right] \, .
    \label{eqn:massThreshold}
\end{equation}

\begin{figure}
    \centering
    \includegraphics[width=0.9\linewidth]{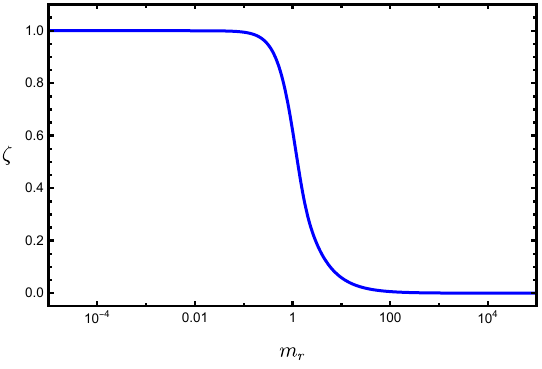}
    \caption{Shown is the threshold profile \cref{eqn:massThreshold} of the mass $\beta$-function \cref{eqn:betaMass}. While $\zeta$ asymptotes to $1$ for small $m_r$, it vanishes at large values. A smooth transitions occurs between these regimes at $m_r \sim 1$.}
    \label{fig:massThreshold}
\end{figure}

In the \ac{UV}, the dimensionless mass $m_r$ becomes small. Expanding \cref{eqn:betaMass} with \cref{eqn:massThreshold} to linear order in $m_r$, we find
\begin{equation}
    \frac{\beta_{m_r}}{m_r} = - 1 - \frac{3 \alpha}{2 \pi} C_F \, ,
    \label{eqn:massRunMS}
\end{equation}
which agrees with the mass anomalous dimension in the \ac{MS} scheme. Due to this anomalous scaling, the dimensionful mass is no longer constant. However, as we have argued already in \cref{fig:qcdRunning}, the change is numerically small. Indeed, assuming $\alpha \sim 0.1$ and $N_c = 3$, we find $\beta_{m_r}/m_r \sim -1.06$. 

In the \ac{IR}, the dimensionless fermion masses grow large. Taking the limit $m_r \to \infty$, \cref{eqn:betaMass} gives rise to
\begin{equation}
    \beta_{m_r} = -m_r - \alpha + \mathcal{O} \left( m_r^{-1} \right) \, .
\end{equation}
Note that the first correction to classical scaling is not proportional to $m_r$. Since $m_r$ grows large for $\mu \to 0$, it follows that, within a perturbative domain, corrections to the classical scaling become increasingly negligible. Thus, far below the threshold the quantum induced anomalous scalings are effectively zero and dimensionful masses become constant. As shown in \cref{fig:massThreshold}, eq.~\cref{eqn:massThreshold} leads to a smooth transition between this \ac{IR} regime and the well-known \ac{UV} regime with effectively massless quarks.

\section{Conclusions}
In this work, we have demonstrated how the poles of \ac{DR} in $d \geq 4$ can be used to define a mass-dependent renormalization scheme. The resulting sensitivity to mass scales allows for a natural implementation of heavy-particle decouplings through smooth thresholds. Compared to other mass-dependent schemes like \ac{MOM}, our approach offers several advantages as it relies solely on pole terms rather than finite parts. The latter often necessitate more demanding computations and induce ambiguities such as gauge dependence. Our scheme avoids this and retains many advantages of \ac{MS}.

To demonstrate the practical applicability of this scheme, we have computed the one-loop \ac{RG} functions in \ac{QCD}. At high energies, standard \ac{MS} results are recovered since quarks become effectively massless. However, near the mass threshold, quark contributions become \textit{smoothly} suppressed, leading to a complete decoupling far below the threshold. Thus, the \ac{RG} running of the corresponding low-energy \ac{EFT} is recovered automatically---heavy degrees of freedom completely decouple and their dimensionful masses become exactly constant.

Several practical applications of our results are conceivable, particularly where finite masses and threshold effects are relevant. One application concerns the \ac{RG} running across multiple energy scales, where several particle thresholds may be encountered. This occurs for example in the evolution of \ac{SM} couplings to the \ac{GUT} scale or the Planck scale \cite{Degrassi:2012ry}. In standard approaches based on \ac{MS}, this requires a sequence of matching conditions to connect low-energy \acp{EFT} to their high-energy completion \cite{Burgess:2007pt}. Since the choice of matching scale is inherently arbitrary, the resulting scale setting ambiguity induces significant uncertainties. Our renormalization scheme avoids such issues since it naturally implements smooth thresholds and matching conditions can be avoided entirely. Beyond that, our scheme could also be useful for scattering processes at threshold energies. By directly integrating threshold effects into \ac{RG} evolutions, our scheme could be beneficial for the resummation of large logarithms \cite{Beneke:2011mq}. Other areas where this scheme could be useful due to the relevance of finite quark masses include computations in quark matter \cite{Kurkela:2009gj}, or the variable flavour number scheme \cite{Thorne:2006qt}.

Regarding future developments, a natural next step would be the extension to higher loop orders. While finite parts are avoided in our scheme, computing poles in $d > 4$ turns this process more demanding than \ac{MS}. However, relevant information on threshold effects could potentially already be obtained by computing finitely many poles in $d > 4$. On a different tack, it should be emphasized that the renormalization scheme used here is not restricted to \ac{DR}. Since poles in \ac{DR} correspond to logarithmic divergences, this approach could be adapted to any other regularization by subtracting all logarithmic divergences obtained in $d \geq 4$.  Furthermore, the scheme invariance of logarithmic divergences (up to two-loop) means that results obtained this way should be independent of the specific regularization that was used.

\section*{Acknowledgments}
I would like to thank Bob Holdom for inspiring discussions on threshold effects, Charlie Cresswell-Hogg and Daniel Litim for discussions on related topics, and Robert Harlander for comments on an earlier version of this manuscript. I acknowledge usage of Google Gemini 3.1 Pro for assistance with language editing as well as literature research. This work was supported by the Deutsche Forschungsgemeinschaft (DFG) through the Walter-Benjamin program (KL 3849/1-1, project no. 562668270).

\bibliographystyle{apsrev4-2}
\bibliography{references}

\begin{thebibliography}{52}%
\makeatletter
\providecommand \@ifxundefined [1]{%
 \@ifx{#1\undefined}
}%
\providecommand \@ifnum [1]{%
 \ifnum #1\expandafter \@firstoftwo
 \else \expandafter \@secondoftwo
 \fi
}%
\providecommand \@ifx [1]{%
 \ifx #1\expandafter \@firstoftwo
 \else \expandafter \@secondoftwo
 \fi
}%
\providecommand \natexlab [1]{#1}%
\providecommand \enquote  [1]{``#1''}%
\providecommand \bibnamefont  [1]{#1}%
\providecommand \bibfnamefont [1]{#1}%
\providecommand \citenamefont [1]{#1}%
\providecommand \href@noop [0]{\@secondoftwo}%
\providecommand \href [0]{\begingroup \@sanitize@url \@href}%
\providecommand \@href[1]{\@@startlink{#1}\@@href}%
\providecommand \@@href[1]{\endgroup#1\@@endlink}%
\providecommand \@sanitize@url [0]{\catcode `\\12\catcode `\$12\catcode `\&12\catcode `\#12\catcode `\^12\catcode `\_12\catcode `\%12\relax}%
\providecommand \@@startlink[1]{}%
\providecommand \@@endlink[0]{}%
\providecommand \url  [0]{\begingroup\@sanitize@url \@url }%
\providecommand \@url [1]{\endgroup\@href {#1}{\urlprefix }}%
\providecommand \urlprefix  [0]{URL }%
\providecommand \Eprint [0]{\href }%
\providecommand \doibase [0]{https://doi.org/}%
\providecommand \selectlanguage [0]{\@gobble}%
\providecommand \bibinfo  [0]{\@secondoftwo}%
\providecommand \bibfield  [0]{\@secondoftwo}%
\providecommand \translation [1]{[#1]}%
\providecommand \BibitemOpen [0]{}%
\providecommand \bibitemStop [0]{}%
\providecommand \bibitemNoStop [0]{.\EOS\space}%
\providecommand \EOS [0]{\spacefactor3000\relax}%
\providecommand \BibitemShut  [1]{\csname bibitem#1\endcsname}%
\let\auto@bib@innerbib\@empty
%</preamble>
\bibitem [{\citenamefont {Appelquist}\ and\ \citenamefont {Carazzone}(1975)}]{Appelquist:1974tg}%
  \BibitemOpen
  \bibfield  {author} {\bibinfo {author} {\bibfnamefont {T.}~\bibnamefont {Appelquist}}\ and\ \bibinfo {author} {\bibfnamefont {J.}~\bibnamefont {Carazzone}},\ }\href {https://doi.org/10.1103/PhysRevD.11.2856} {\bibfield  {journal} {\bibinfo  {journal} {Phys. Rev. D}\ }\textbf {\bibinfo {volume} {11}},\ \bibinfo {pages} {2856} (\bibinfo {year} {1975})}\BibitemShut {NoStop}%
\bibitem [{\citenamefont {van Ritbergen}\ \emph {et~al.}(1997)\citenamefont {van Ritbergen}, \citenamefont {Vermaseren},\ and\ \citenamefont {Larin}}]{vanRitbergen:1997va}%
  \BibitemOpen
  \bibfield  {author} {\bibinfo {author} {\bibfnamefont {T.}~\bibnamefont {van Ritbergen}}, \bibinfo {author} {\bibfnamefont {J.~A.~M.}\ \bibnamefont {Vermaseren}},\ and\ \bibinfo {author} {\bibfnamefont {S.~A.}\ \bibnamefont {Larin}},\ }\href {https://doi.org/10.1016/S0370-2693(97)00370-5} {\bibfield  {journal} {\bibinfo  {journal} {Phys. Lett. B}\ }\textbf {\bibinfo {volume} {400}},\ \bibinfo {pages} {379} (\bibinfo {year} {1997})},\ \Eprint {https://arxiv.org/abs/hep-ph/9701390} {arXiv:hep-ph/9701390} \BibitemShut {NoStop}%
\bibitem [{\citenamefont {Czakon}(2005)}]{Czakon:2004bu}%
  \BibitemOpen
  \bibfield  {author} {\bibinfo {author} {\bibfnamefont {M.}~\bibnamefont {Czakon}},\ }\href {https://doi.org/10.1016/j.nuclphysb.2005.01.012} {\bibfield  {journal} {\bibinfo  {journal} {Nucl. Phys. B}\ }\textbf {\bibinfo {volume} {710}},\ \bibinfo {pages} {485} (\bibinfo {year} {2005})},\ \Eprint {https://arxiv.org/abs/hep-ph/0411261} {arXiv:hep-ph/0411261} \BibitemShut {NoStop}%
\bibitem [{\citenamefont {Baikov}\ \emph {et~al.}(2017)\citenamefont {Baikov}, \citenamefont {Chetyrkin},\ and\ \citenamefont {K{\"u}hn}}]{Baikov:2016tgj}%
  \BibitemOpen
  \bibfield  {author} {\bibinfo {author} {\bibfnamefont {P.~A.}\ \bibnamefont {Baikov}}, \bibinfo {author} {\bibfnamefont {K.~G.}\ \bibnamefont {Chetyrkin}},\ and\ \bibinfo {author} {\bibfnamefont {J.~H.}\ \bibnamefont {K{\"u}hn}},\ }\href {https://doi.org/10.1103/PhysRevLett.118.082002} {\bibfield  {journal} {\bibinfo  {journal} {Phys. Rev. Lett.}\ }\textbf {\bibinfo {volume} {118}},\ \bibinfo {pages} {082002} (\bibinfo {year} {2017})},\ \Eprint {https://arxiv.org/abs/1606.08659} {arXiv:1606.08659 [hep-ph]} \BibitemShut {NoStop}%
\bibitem [{\citenamefont {Schnetz}(2018)}]{Schnetz:2016fhy}%
  \BibitemOpen
  \bibfield  {author} {\bibinfo {author} {\bibfnamefont {O.}~\bibnamefont {Schnetz}},\ }\href {https://doi.org/10.1103/PhysRevD.97.085018} {\bibfield  {journal} {\bibinfo  {journal} {Phys. Rev. D}\ }\textbf {\bibinfo {volume} {97}},\ \bibinfo {pages} {085018} (\bibinfo {year} {2018})},\ \Eprint {https://arxiv.org/abs/1606.08598} {arXiv:1606.08598 [hep-th]} \BibitemShut {NoStop}%
\bibitem [{\citenamefont {Herzog}\ \emph {et~al.}(2017)\citenamefont {Herzog}, \citenamefont {Ruijl}, \citenamefont {Ueda}, \citenamefont {Vermaseren},\ and\ \citenamefont {Vogt}}]{Herzog:2017ohr}%
  \BibitemOpen
  \bibfield  {author} {\bibinfo {author} {\bibfnamefont {F.}~\bibnamefont {Herzog}}, \bibinfo {author} {\bibfnamefont {B.}~\bibnamefont {Ruijl}}, \bibinfo {author} {\bibfnamefont {T.}~\bibnamefont {Ueda}}, \bibinfo {author} {\bibfnamefont {J.~A.~M.}\ \bibnamefont {Vermaseren}},\ and\ \bibinfo {author} {\bibfnamefont {A.}~\bibnamefont {Vogt}},\ }\href {https://doi.org/10.1007/JHEP02(2017)090} {\bibfield  {journal} {\bibinfo  {journal} {JHEP}\ }\textbf {\bibinfo {volume} {02}},\ \bibinfo {pages} {090}},\ \Eprint {https://arxiv.org/abs/1701.01404} {arXiv:1701.01404 [hep-ph]} \BibitemShut {NoStop}%
\bibitem [{\citenamefont {Bernreuther}(1994)}]{Bernreuther:1994nw}%
  \BibitemOpen
  \bibfield  {author} {\bibinfo {author} {\bibfnamefont {W.}~\bibnamefont {Bernreuther}},\ }in\ \href@noop {} {\emph {\bibinfo {booktitle} {{Workshop on QCD at LEP}}}}\ (\bibinfo {year} {1994})\ pp.\ \bibinfo {pages} {39--46},\ \Eprint {https://arxiv.org/abs/hep-ph/9409390} {arXiv:hep-ph/9409390} \BibitemShut {NoStop}%
\bibitem [{\citenamefont {Beneke}\ \emph {et~al.}(2012)\citenamefont {Beneke}, \citenamefont {Falgari}, \citenamefont {Klein},\ and\ \citenamefont {Schwinn}}]{Beneke:2011mq}%
  \BibitemOpen
  \bibfield  {author} {\bibinfo {author} {\bibfnamefont {M.}~\bibnamefont {Beneke}}, \bibinfo {author} {\bibfnamefont {P.}~\bibnamefont {Falgari}}, \bibinfo {author} {\bibfnamefont {S.}~\bibnamefont {Klein}},\ and\ \bibinfo {author} {\bibfnamefont {C.}~\bibnamefont {Schwinn}},\ }\href {https://doi.org/10.1016/j.nuclphysb.2011.10.021} {\bibfield  {journal} {\bibinfo  {journal} {Nucl. Phys. B}\ }\textbf {\bibinfo {volume} {855}},\ \bibinfo {pages} {695} (\bibinfo {year} {2012})},\ \Eprint {https://arxiv.org/abs/1109.1536} {arXiv:1109.1536 [hep-ph]} \BibitemShut {NoStop}%
\bibitem [{\citenamefont {Burgess}(2007)}]{Burgess:2007pt}%
  \BibitemOpen
  \bibfield  {author} {\bibinfo {author} {\bibfnamefont {C.~P.}\ \bibnamefont {Burgess}},\ }\href {https://doi.org/10.1146/annurev.nucl.56.080805.140508} {\bibfield  {journal} {\bibinfo  {journal} {Ann. Rev. Nucl. Part. Sci.}\ }\textbf {\bibinfo {volume} {57}},\ \bibinfo {pages} {329} (\bibinfo {year} {2007})},\ \Eprint {https://arxiv.org/abs/hep-th/0701053} {arXiv:hep-th/0701053} \BibitemShut {NoStop}%
\bibitem [{\citenamefont {Bernreuther}\ and\ \citenamefont {Wetzel}(1982)}]{Bernreuther:1981sg}%
  \BibitemOpen
  \bibfield  {author} {\bibinfo {author} {\bibfnamefont {W.}~\bibnamefont {Bernreuther}}\ and\ \bibinfo {author} {\bibfnamefont {W.}~\bibnamefont {Wetzel}},\ }\href {https://doi.org/10.1016/0550-3213(82)90288-7} {\bibfield  {journal} {\bibinfo  {journal} {Nucl. Phys. B}\ }\textbf {\bibinfo {volume} {197}},\ \bibinfo {pages} {228} (\bibinfo {year} {1982})},\ \bibinfo {note} {[Erratum: Nucl.Phys.B 513, 758--758 (1998)]}\BibitemShut {NoStop}%
\bibitem [{\citenamefont {Chetyrkin}\ \emph {et~al.}(1998{\natexlab{a}})\citenamefont {Chetyrkin}, \citenamefont {Kniehl},\ and\ \citenamefont {Steinhauser}}]{Chetyrkin:1997un}%
  \BibitemOpen
  \bibfield  {author} {\bibinfo {author} {\bibfnamefont {K.~G.}\ \bibnamefont {Chetyrkin}}, \bibinfo {author} {\bibfnamefont {B.~A.}\ \bibnamefont {Kniehl}},\ and\ \bibinfo {author} {\bibfnamefont {M.}~\bibnamefont {Steinhauser}},\ }\href {https://doi.org/10.1016/S0550-3213(97)00649-4} {\bibfield  {journal} {\bibinfo  {journal} {Nucl. Phys. B}\ }\textbf {\bibinfo {volume} {510}},\ \bibinfo {pages} {61} (\bibinfo {year} {1998}{\natexlab{a}})},\ \Eprint {https://arxiv.org/abs/hep-ph/9708255} {arXiv:hep-ph/9708255} \BibitemShut {NoStop}%
\bibitem [{\citenamefont {Chetyrkin}\ \emph {et~al.}(2000)\citenamefont {Chetyrkin}, \citenamefont {Kuhn},\ and\ \citenamefont {Steinhauser}}]{Chetyrkin:2000yt}%
  \BibitemOpen
  \bibfield  {author} {\bibinfo {author} {\bibfnamefont {K.~G.}\ \bibnamefont {Chetyrkin}}, \bibinfo {author} {\bibfnamefont {J.~H.}\ \bibnamefont {Kuhn}},\ and\ \bibinfo {author} {\bibfnamefont {M.}~\bibnamefont {Steinhauser}},\ }\href {https://doi.org/10.1016/S0010-4655(00)00155-7} {\bibfield  {journal} {\bibinfo  {journal} {Comput. Phys. Commun.}\ }\textbf {\bibinfo {volume} {133}},\ \bibinfo {pages} {43} (\bibinfo {year} {2000})},\ \Eprint {https://arxiv.org/abs/hep-ph/0004189} {arXiv:hep-ph/0004189} \BibitemShut {NoStop}%
\bibitem [{\citenamefont {Chetyrkin}\ \emph {et~al.}(2006)\citenamefont {Chetyrkin}, \citenamefont {Kuhn},\ and\ \citenamefont {Sturm}}]{Chetyrkin:2005ia}%
  \BibitemOpen
  \bibfield  {author} {\bibinfo {author} {\bibfnamefont {K.~G.}\ \bibnamefont {Chetyrkin}}, \bibinfo {author} {\bibfnamefont {J.~H.}\ \bibnamefont {Kuhn}},\ and\ \bibinfo {author} {\bibfnamefont {C.}~\bibnamefont {Sturm}},\ }\href {https://doi.org/10.1016/j.nuclphysb.2006.03.020} {\bibfield  {journal} {\bibinfo  {journal} {Nucl. Phys. B}\ }\textbf {\bibinfo {volume} {744}},\ \bibinfo {pages} {121} (\bibinfo {year} {2006})},\ \Eprint {https://arxiv.org/abs/hep-ph/0512060} {arXiv:hep-ph/0512060} \BibitemShut {NoStop}%
\bibitem [{\citenamefont {Schroder}\ and\ \citenamefont {Steinhauser}(2006)}]{Schroder:2005hy}%
  \BibitemOpen
  \bibfield  {author} {\bibinfo {author} {\bibfnamefont {Y.}~\bibnamefont {Schroder}}\ and\ \bibinfo {author} {\bibfnamefont {M.}~\bibnamefont {Steinhauser}},\ }\href {https://doi.org/10.1088/1126-6708/2006/01/051} {\bibfield  {journal} {\bibinfo  {journal} {JHEP}\ }\textbf {\bibinfo {volume} {01}},\ \bibinfo {pages} {051}},\ \Eprint {https://arxiv.org/abs/hep-ph/0512058} {arXiv:hep-ph/0512058} \BibitemShut {NoStop}%
\bibitem [{\citenamefont {Brodsky}\ \emph {et~al.}(1998)\citenamefont {Brodsky}, \citenamefont {Gill}, \citenamefont {Melles},\ and\ \citenamefont {Rathsman}}]{Brodsky:1998mf}%
  \BibitemOpen
  \bibfield  {author} {\bibinfo {author} {\bibfnamefont {S.~J.}\ \bibnamefont {Brodsky}}, \bibinfo {author} {\bibfnamefont {M.~S.}\ \bibnamefont {Gill}}, \bibinfo {author} {\bibfnamefont {M.}~\bibnamefont {Melles}},\ and\ \bibinfo {author} {\bibfnamefont {J.}~\bibnamefont {Rathsman}},\ }\href {https://doi.org/10.1103/PhysRevD.58.116006} {\bibfield  {journal} {\bibinfo  {journal} {Phys. Rev. D}\ }\textbf {\bibinfo {volume} {58}},\ \bibinfo {pages} {116006} (\bibinfo {year} {1998})},\ \Eprint {https://arxiv.org/abs/hep-ph/9801330} {arXiv:hep-ph/9801330} \BibitemShut {NoStop}%
\bibitem [{\citenamefont {Georgi}\ and\ \citenamefont {Politzer}(1976)}]{Georgi:1976ve}%
  \BibitemOpen
  \bibfield  {author} {\bibinfo {author} {\bibfnamefont {H.}~\bibnamefont {Georgi}}\ and\ \bibinfo {author} {\bibfnamefont {H.~D.}\ \bibnamefont {Politzer}},\ }\href {https://doi.org/10.1103/PhysRevD.14.1829} {\bibfield  {journal} {\bibinfo  {journal} {Phys. Rev. D}\ }\textbf {\bibinfo {volume} {14}},\ \bibinfo {pages} {1829} (\bibinfo {year} {1976})}\BibitemShut {NoStop}%
\bibitem [{\citenamefont {Celmaster}\ and\ \citenamefont {Gonsalves}(1979)}]{Celmaster:1979km}%
  \BibitemOpen
  \bibfield  {author} {\bibinfo {author} {\bibfnamefont {W.}~\bibnamefont {Celmaster}}\ and\ \bibinfo {author} {\bibfnamefont {R.~J.}\ \bibnamefont {Gonsalves}},\ }\href {https://doi.org/10.1103/PhysRevD.20.1420} {\bibfield  {journal} {\bibinfo  {journal} {Phys. Rev. D}\ }\textbf {\bibinfo {volume} {20}},\ \bibinfo {pages} {1420} (\bibinfo {year} {1979})}\BibitemShut {NoStop}%
\bibitem [{\citenamefont {Chetyrkin}\ and\ \citenamefont {Seidensticker}(2000)}]{Chetyrkin:2000fd}%
  \BibitemOpen
  \bibfield  {author} {\bibinfo {author} {\bibfnamefont {K.~G.}\ \bibnamefont {Chetyrkin}}\ and\ \bibinfo {author} {\bibfnamefont {T.}~\bibnamefont {Seidensticker}},\ }\href {https://doi.org/10.1016/S0370-2693(00)01217-X} {\bibfield  {journal} {\bibinfo  {journal} {Phys. Lett. B}\ }\textbf {\bibinfo {volume} {495}},\ \bibinfo {pages} {74} (\bibinfo {year} {2000})},\ \Eprint {https://arxiv.org/abs/hep-ph/0008094} {arXiv:hep-ph/0008094} \BibitemShut {NoStop}%
\bibitem [{\citenamefont {Gracey}(2011{\natexlab{a}})}]{Gracey:2011vw}%
  \BibitemOpen
  \bibfield  {author} {\bibinfo {author} {\bibfnamefont {J.~A.}\ \bibnamefont {Gracey}},\ }\href {https://doi.org/10.1103/PhysRevD.84.085011} {\bibfield  {journal} {\bibinfo  {journal} {Phys. Rev. D}\ }\textbf {\bibinfo {volume} {84}},\ \bibinfo {pages} {085011} (\bibinfo {year} {2011}{\natexlab{a}})},\ \Eprint {https://arxiv.org/abs/1108.4806} {arXiv:1108.4806 [hep-ph]} \BibitemShut {NoStop}%
\bibitem [{\citenamefont {Gracey}(2011{\natexlab{b}})}]{Gracey:2011pf}%
  \BibitemOpen
  \bibfield  {author} {\bibinfo {author} {\bibfnamefont {J.~A.}\ \bibnamefont {Gracey}},\ }\href {https://doi.org/10.1016/j.physletb.2011.04.052} {\bibfield  {journal} {\bibinfo  {journal} {Phys. Lett. B}\ }\textbf {\bibinfo {volume} {700}},\ \bibinfo {pages} {79} (\bibinfo {year} {2011}{\natexlab{b}})},\ \Eprint {https://arxiv.org/abs/1104.5382} {arXiv:1104.5382 [hep-ph]} \BibitemShut {NoStop}%
\bibitem [{\citenamefont {Bednyakov}\ and\ \citenamefont {Pikelner}(2020)}]{Bednyakov:2020cdf}%
  \BibitemOpen
  \bibfield  {author} {\bibinfo {author} {\bibfnamefont {A.}~\bibnamefont {Bednyakov}}\ and\ \bibinfo {author} {\bibfnamefont {A.}~\bibnamefont {Pikelner}},\ }\href {https://doi.org/10.1103/PhysRevD.101.071502} {\bibfield  {journal} {\bibinfo  {journal} {Phys. Rev. D}\ }\textbf {\bibinfo {volume} {101}},\ \bibinfo {pages} {071502} (\bibinfo {year} {2020})},\ \Eprint {https://arxiv.org/abs/2002.02875} {arXiv:2002.02875 [hep-ph]} \BibitemShut {NoStop}%
\bibitem [{\citenamefont {Abbott}(1981)}]{Abbott:1980hw}%
  \BibitemOpen
  \bibfield  {author} {\bibinfo {author} {\bibfnamefont {L.~F.}\ \bibnamefont {Abbott}},\ }\href {https://doi.org/10.1016/0550-3213(81)90371-0} {\bibfield  {journal} {\bibinfo  {journal} {Nucl. Phys. B}\ }\textbf {\bibinfo {volume} {185}},\ \bibinfo {pages} {189} (\bibinfo {year} {1981})}\BibitemShut {NoStop}%
\bibitem [{\citenamefont {Rebhan}(1986)}]{Rebhan:1985yf}%
  \BibitemOpen
  \bibfield  {author} {\bibinfo {author} {\bibfnamefont {A.}~\bibnamefont {Rebhan}},\ }\href {https://doi.org/10.1007/BF01575440} {\bibfield  {journal} {\bibinfo  {journal} {Z. Phys. C}\ }\textbf {\bibinfo {volume} {30}},\ \bibinfo {pages} {309} (\bibinfo {year} {1986})}\BibitemShut {NoStop}%
\bibitem [{\citenamefont {Chetyrkin}\ \emph {et~al.}(2009)\citenamefont {Chetyrkin}, \citenamefont {Kniehl},\ and\ \citenamefont {Steinhauser}}]{Chetyrkin:2008jk}%
  \BibitemOpen
  \bibfield  {author} {\bibinfo {author} {\bibfnamefont {K.~G.}\ \bibnamefont {Chetyrkin}}, \bibinfo {author} {\bibfnamefont {B.~A.}\ \bibnamefont {Kniehl}},\ and\ \bibinfo {author} {\bibfnamefont {M.}~\bibnamefont {Steinhauser}},\ }\href {https://doi.org/10.1016/j.nuclphysb.2009.01.026} {\bibfield  {journal} {\bibinfo  {journal} {Nucl. Phys. B}\ }\textbf {\bibinfo {volume} {814}},\ \bibinfo {pages} {231} (\bibinfo {year} {2009})},\ \Eprint {https://arxiv.org/abs/0812.1337} {arXiv:0812.1337 [hep-ph]} \BibitemShut {NoStop}%
\bibitem [{\citenamefont {Jegerlehner}\ and\ \citenamefont {Tarasov}(1999)}]{Jegerlehner:1998zg}%
  \BibitemOpen
  \bibfield  {author} {\bibinfo {author} {\bibfnamefont {F.}~\bibnamefont {Jegerlehner}}\ and\ \bibinfo {author} {\bibfnamefont {O.~V.}\ \bibnamefont {Tarasov}},\ }\href {https://doi.org/10.1016/S0550-3213(99)00141-8} {\bibfield  {journal} {\bibinfo  {journal} {Nucl. Phys. B}\ }\textbf {\bibinfo {volume} {549}},\ \bibinfo {pages} {481} (\bibinfo {year} {1999})},\ \Eprint {https://arxiv.org/abs/hep-ph/9809485} {arXiv:hep-ph/9809485} \BibitemShut {NoStop}%
\bibitem [{\citenamefont {Zeng}\ \emph {et~al.}(2020)\citenamefont {Zeng}, \citenamefont {Wu}, \citenamefont {Zheng},\ and\ \citenamefont {Shen}}]{Zeng:2020lwi}%
  \BibitemOpen
  \bibfield  {author} {\bibinfo {author} {\bibfnamefont {J.}~\bibnamefont {Zeng}}, \bibinfo {author} {\bibfnamefont {X.-G.}\ \bibnamefont {Wu}}, \bibinfo {author} {\bibfnamefont {X.-C.}\ \bibnamefont {Zheng}},\ and\ \bibinfo {author} {\bibfnamefont {J.-M.}\ \bibnamefont {Shen}},\ }\href {https://doi.org/10.1088/1674-1137/abae4e} {\bibfield  {journal} {\bibinfo  {journal} {Chin. Phys. C}\ }\textbf {\bibinfo {volume} {44}},\ \bibinfo {pages} {113102} (\bibinfo {year} {2020})},\ \Eprint {https://arxiv.org/abs/2004.12068} {arXiv:2004.12068 [hep-ph]} \BibitemShut {NoStop}%
\bibitem [{\citenamefont {Veltman}(1981)}]{Veltman:1980mj}%
  \BibitemOpen
  \bibfield  {author} {\bibinfo {author} {\bibfnamefont {M.~J.~G.}\ \bibnamefont {Veltman}},\ }\href@noop {} {\bibfield  {journal} {\bibinfo  {journal} {Acta Phys. Polon. B}\ }\textbf {\bibinfo {volume} {12}},\ \bibinfo {pages} {437} (\bibinfo {year} {1981})}\BibitemShut {NoStop}%
\bibitem [{\citenamefont {Castellani}\ and\ \citenamefont {Van~Nieuwenhuizen}(1983)}]{Castellani:1982dk}%
  \BibitemOpen
  \bibfield  {author} {\bibinfo {author} {\bibfnamefont {L.}~\bibnamefont {Castellani}}\ and\ \bibinfo {author} {\bibfnamefont {P.}~\bibnamefont {Van~Nieuwenhuizen}},\ }\href {https://doi.org/10.1016/0550-3213(83)90514-X} {\bibfield  {journal} {\bibinfo  {journal} {Nucl. Phys. B}\ }\textbf {\bibinfo {volume} {213}},\ \bibinfo {pages} {305} (\bibinfo {year} {1983})}\BibitemShut {NoStop}%
\bibitem [{\citenamefont {Jack}\ and\ \citenamefont {Jones}(1990)}]{Jack:1990pz}%
  \BibitemOpen
  \bibfield  {author} {\bibinfo {author} {\bibfnamefont {I.}~\bibnamefont {Jack}}\ and\ \bibinfo {author} {\bibfnamefont {D.~R.~T.}\ \bibnamefont {Jones}},\ }\href {https://doi.org/10.1016/0550-3213(90)90574-W} {\bibfield  {journal} {\bibinfo  {journal} {Nucl. Phys. B}\ }\textbf {\bibinfo {volume} {342}},\ \bibinfo {pages} {127} (\bibinfo {year} {1990})}\BibitemShut {NoStop}%
\bibitem [{\citenamefont {Al-Sarhi}\ \emph {et~al.}(1990)\citenamefont {Al-Sarhi}, \citenamefont {Jones},\ and\ \citenamefont {Jack}}]{Al-Sarhi:1990nmv}%
  \BibitemOpen
  \bibfield  {author} {\bibinfo {author} {\bibfnamefont {M.~S.}\ \bibnamefont {Al-Sarhi}}, \bibinfo {author} {\bibfnamefont {D.~R.~T.}\ \bibnamefont {Jones}},\ and\ \bibinfo {author} {\bibfnamefont {I.}~\bibnamefont {Jack}},\ }\href {https://doi.org/10.1016/0550-3213(90)90394-S} {\bibfield  {journal} {\bibinfo  {journal} {Nucl. Phys. B}\ }\textbf {\bibinfo {volume} {345}},\ \bibinfo {pages} {431} (\bibinfo {year} {1990})}\BibitemShut {NoStop}%
\bibitem [{\citenamefont {Al-sarhi}\ \emph {et~al.}(1992)\citenamefont {Al-sarhi}, \citenamefont {Jack},\ and\ \citenamefont {Jones}}]{Al-sarhi:1991gdi}%
  \BibitemOpen
  \bibfield  {author} {\bibinfo {author} {\bibfnamefont {M.~S.}\ \bibnamefont {Al-sarhi}}, \bibinfo {author} {\bibfnamefont {I.}~\bibnamefont {Jack}},\ and\ \bibinfo {author} {\bibfnamefont {D.~R.~T.}\ \bibnamefont {Jones}},\ }\href {https://doi.org/10.1007/BF01482591} {\bibfield  {journal} {\bibinfo  {journal} {Z. Phys. C}\ }\textbf {\bibinfo {volume} {55}},\ \bibinfo {pages} {283} (\bibinfo {year} {1992})}\BibitemShut {NoStop}%
\bibitem [{\citenamefont {Kaplan}\ \emph {et~al.}(1998)\citenamefont {Kaplan}, \citenamefont {Savage},\ and\ \citenamefont {Wise}}]{Kaplan:1998tg}%
  \BibitemOpen
  \bibfield  {author} {\bibinfo {author} {\bibfnamefont {D.~B.}\ \bibnamefont {Kaplan}}, \bibinfo {author} {\bibfnamefont {M.~J.}\ \bibnamefont {Savage}},\ and\ \bibinfo {author} {\bibfnamefont {M.~B.}\ \bibnamefont {Wise}},\ }\href {https://doi.org/10.1016/S0370-2693(98)00210-X} {\bibfield  {journal} {\bibinfo  {journal} {Phys. Lett. B}\ }\textbf {\bibinfo {volume} {424}},\ \bibinfo {pages} {390} (\bibinfo {year} {1998})},\ \Eprint {https://arxiv.org/abs/nucl-th/9801034} {arXiv:nucl-th/9801034} \BibitemShut {NoStop}%
\bibitem [{\citenamefont {Kluth}(2025)}]{Kluth:2024lar}%
  \BibitemOpen
  \bibfield  {author} {\bibinfo {author} {\bibfnamefont {Y.}~\bibnamefont {Kluth}},\ }\href {https://doi.org/10.1103/PhysRevD.111.106010} {\bibfield  {journal} {\bibinfo  {journal} {Phys. Rev. D}\ }\textbf {\bibinfo {volume} {111}},\ \bibinfo {pages} {106010} (\bibinfo {year} {2025})},\ \Eprint {https://arxiv.org/abs/2409.09252} {arXiv:2409.09252 [hep-th]} \BibitemShut {NoStop}%
\bibitem [{\citenamefont {Cresswell-Hogg}\ \emph {et~al.}(2026)\citenamefont {Cresswell-Hogg}, \citenamefont {Kluth},\ and\ \citenamefont {Litim}}]{cckl}%
  \BibitemOpen
  \bibfield  {author} {\bibinfo {author} {\bibfnamefont {C.}~\bibnamefont {Cresswell-Hogg}}, \bibinfo {author} {\bibfnamefont {Y.}~\bibnamefont {Kluth}},\ and\ \bibinfo {author} {\bibfnamefont {D.~F.}\ \bibnamefont {Litim}}} (\bibinfo {year} {2026}),\ \bibinfo {note} {to appear}\BibitemShut {NoStop}%
\bibitem [{\citenamefont {Weinberg}(1980)}]{Weinberg:1980gg}%
  \BibitemOpen
  \bibfield  {author} {\bibinfo {author} {\bibfnamefont {S.}~\bibnamefont {Weinberg}},\ }\bibinfo {title} {{Ultraviolet divergences in quantum theories of gravitation}},\ in\ \href@noop {} {\emph {\bibinfo {booktitle} {{General Relativity}: {An Einstein Centenary Survey}}}}\ (\bibinfo {year} {1980})\ pp.\ \bibinfo {pages} {790--831}\BibitemShut {NoStop}%
\bibitem [{\citenamefont {Falls}\ and\ \citenamefont {Ferrero}(2025)}]{Falls:2024noj}%
  \BibitemOpen
  \bibfield  {author} {\bibinfo {author} {\bibfnamefont {K.}~\bibnamefont {Falls}}\ and\ \bibinfo {author} {\bibfnamefont {R.}~\bibnamefont {Ferrero}},\ }\href {https://doi.org/10.1007/JHEP08(2025)173} {\bibfield  {journal} {\bibinfo  {journal} {JHEP}\ }\textbf {\bibinfo {volume} {08}},\ \bibinfo {pages} {173}},\ \Eprint {https://arxiv.org/abs/2411.00938} {arXiv:2411.00938 [hep-th]} \BibitemShut {NoStop}%
\bibitem [{\citenamefont {Wilson}\ and\ \citenamefont {Fisher}(1972)}]{Wilson:1971dc}%
  \BibitemOpen
  \bibfield  {author} {\bibinfo {author} {\bibfnamefont {K.~G.}\ \bibnamefont {Wilson}}\ and\ \bibinfo {author} {\bibfnamefont {M.~E.}\ \bibnamefont {Fisher}},\ }\href {https://doi.org/10.1103/PhysRevLett.28.240} {\bibfield  {journal} {\bibinfo  {journal} {Phys. Rev. Lett.}\ }\textbf {\bibinfo {volume} {28}},\ \bibinfo {pages} {240} (\bibinfo {year} {1972})}\BibitemShut {NoStop}%
\bibitem [{\citenamefont {Guida}\ and\ \citenamefont {Zinn-Justin}(1998)}]{Guida:1998bx}%
  \BibitemOpen
  \bibfield  {author} {\bibinfo {author} {\bibfnamefont {R.}~\bibnamefont {Guida}}\ and\ \bibinfo {author} {\bibfnamefont {J.}~\bibnamefont {Zinn-Justin}},\ }\href {https://doi.org/10.1088/0305-4470/31/40/006} {\bibfield  {journal} {\bibinfo  {journal} {J. Phys. A}\ }\textbf {\bibinfo {volume} {31}},\ \bibinfo {pages} {8103} (\bibinfo {year} {1998})},\ \Eprint {https://arxiv.org/abs/cond-mat/9803240} {arXiv:cond-mat/9803240} \BibitemShut {NoStop}%
\bibitem [{\citenamefont {Hasenbusch}(2010)}]{Hasenbusch:2010hkh}%
  \BibitemOpen
  \bibfield  {author} {\bibinfo {author} {\bibfnamefont {M.}~\bibnamefont {Hasenbusch}},\ }\href {https://doi.org/10.1103/PhysRevB.82.174433} {\bibfield  {journal} {\bibinfo  {journal} {Phys. Rev. B}\ }\textbf {\bibinfo {volume} {82}},\ \bibinfo {pages} {174433} (\bibinfo {year} {2010})},\ \Eprint {https://arxiv.org/abs/1004.4486} {arXiv:1004.4486 [cond-mat.stat-mech]} \BibitemShut {NoStop}%
\bibitem [{\citenamefont {El-Showk}\ \emph {et~al.}(2014)\citenamefont {El-Showk}, \citenamefont {Paulos}, \citenamefont {Poland}, \citenamefont {Rychkov}, \citenamefont {Simmons-Duffin},\ and\ \citenamefont {Vichi}}]{El-Showk:2014dwa}%
  \BibitemOpen
  \bibfield  {author} {\bibinfo {author} {\bibfnamefont {S.}~\bibnamefont {El-Showk}}, \bibinfo {author} {\bibfnamefont {M.~F.}\ \bibnamefont {Paulos}}, \bibinfo {author} {\bibfnamefont {D.}~\bibnamefont {Poland}}, \bibinfo {author} {\bibfnamefont {S.}~\bibnamefont {Rychkov}}, \bibinfo {author} {\bibfnamefont {D.}~\bibnamefont {Simmons-Duffin}},\ and\ \bibinfo {author} {\bibfnamefont {A.}~\bibnamefont {Vichi}},\ }\href {https://doi.org/10.1007/s10955-014-1042-7} {\bibfield  {journal} {\bibinfo  {journal} {J. Stat. Phys.}\ }\textbf {\bibinfo {volume} {157}},\ \bibinfo {pages} {869} (\bibinfo {year} {2014})},\ \Eprint {https://arxiv.org/abs/1403.4545} {arXiv:1403.4545 [hep-th]} \BibitemShut {NoStop}%
\bibitem [{\citenamefont {Kos}\ \emph {et~al.}(2016)\citenamefont {Kos}, \citenamefont {Poland}, \citenamefont {Simmons-Duffin},\ and\ \citenamefont {Vichi}}]{Kos:2016ysd}%
  \BibitemOpen
  \bibfield  {author} {\bibinfo {author} {\bibfnamefont {F.}~\bibnamefont {Kos}}, \bibinfo {author} {\bibfnamefont {D.}~\bibnamefont {Poland}}, \bibinfo {author} {\bibfnamefont {D.}~\bibnamefont {Simmons-Duffin}},\ and\ \bibinfo {author} {\bibfnamefont {A.}~\bibnamefont {Vichi}},\ }\href {https://doi.org/10.1007/JHEP08(2016)036} {\bibfield  {journal} {\bibinfo  {journal} {JHEP}\ }\textbf {\bibinfo {volume} {08}},\ \bibinfo {pages} {036}},\ \Eprint {https://arxiv.org/abs/1603.04436} {arXiv:1603.04436 [hep-th]} \BibitemShut {NoStop}%
\bibitem [{\citenamefont {Kompaniets}\ and\ \citenamefont {Panzer}(2017)}]{Kompaniets:2017yct}%
  \BibitemOpen
  \bibfield  {author} {\bibinfo {author} {\bibfnamefont {M.~V.}\ \bibnamefont {Kompaniets}}\ and\ \bibinfo {author} {\bibfnamefont {E.}~\bibnamefont {Panzer}},\ }\href {https://doi.org/10.1103/PhysRevD.96.036016} {\bibfield  {journal} {\bibinfo  {journal} {Phys. Rev. D}\ }\textbf {\bibinfo {volume} {96}},\ \bibinfo {pages} {036016} (\bibinfo {year} {2017})},\ \Eprint {https://arxiv.org/abs/1705.06483} {arXiv:1705.06483 [hep-th]} \BibitemShut {NoStop}%
\bibitem [{\citenamefont {Chang}\ \emph {et~al.}(2025)\citenamefont {Chang}, \citenamefont {Dommes}, \citenamefont {Erramilli}, \citenamefont {Homrich}, \citenamefont {Kravchuk}, \citenamefont {Liu}, \citenamefont {Mitchell}, \citenamefont {Poland},\ and\ \citenamefont {Simmons-Duffin}}]{Chang:2024whx}%
  \BibitemOpen
  \bibfield  {author} {\bibinfo {author} {\bibfnamefont {C.-H.}\ \bibnamefont {Chang}}, \bibinfo {author} {\bibfnamefont {V.}~\bibnamefont {Dommes}}, \bibinfo {author} {\bibfnamefont {R.~S.}\ \bibnamefont {Erramilli}}, \bibinfo {author} {\bibfnamefont {A.}~\bibnamefont {Homrich}}, \bibinfo {author} {\bibfnamefont {P.}~\bibnamefont {Kravchuk}}, \bibinfo {author} {\bibfnamefont {A.}~\bibnamefont {Liu}}, \bibinfo {author} {\bibfnamefont {M.~S.}\ \bibnamefont {Mitchell}}, \bibinfo {author} {\bibfnamefont {D.}~\bibnamefont {Poland}},\ and\ \bibinfo {author} {\bibfnamefont {D.}~\bibnamefont {Simmons-Duffin}},\ }\href {https://doi.org/10.1007/JHEP03(2025)136} {\bibfield  {journal} {\bibinfo  {journal} {JHEP}\ }\textbf {\bibinfo {volume} {03}},\ \bibinfo {pages} {136}},\ \Eprint {https://arxiv.org/abs/2411.15300} {arXiv:2411.15300 [hep-th]} \BibitemShut {NoStop}%
\bibitem [{\citenamefont {Harlander}\ \emph {et~al.}(2020)\citenamefont {Harlander}, \citenamefont {Klein},\ and\ \citenamefont {Lipp}}]{Harlander:2020cyh}%
  \BibitemOpen
  \bibfield  {author} {\bibinfo {author} {\bibfnamefont {R.~V.}\ \bibnamefont {Harlander}}, \bibinfo {author} {\bibfnamefont {S.~Y.}\ \bibnamefont {Klein}},\ and\ \bibinfo {author} {\bibfnamefont {M.}~\bibnamefont {Lipp}},\ }\href {https://doi.org/10.1016/j.cpc.2020.107465} {\bibfield  {journal} {\bibinfo  {journal} {Comput. Phys. Commun.}\ }\textbf {\bibinfo {volume} {256}},\ \bibinfo {pages} {107465} (\bibinfo {year} {2020})},\ \Eprint {https://arxiv.org/abs/2003.00896} {arXiv:2003.00896 [physics.ed-ph]} \BibitemShut {NoStop}%
\bibitem [{\citenamefont {Harlander}\ \emph {et~al.}(2024)\citenamefont {Harlander}, \citenamefont {Klein},\ and\ \citenamefont {Schaaf}}]{Harlander:2024qbn}%
  \BibitemOpen
  \bibfield  {author} {\bibinfo {author} {\bibfnamefont {R.}~\bibnamefont {Harlander}}, \bibinfo {author} {\bibfnamefont {S.~Y.}\ \bibnamefont {Klein}},\ and\ \bibinfo {author} {\bibfnamefont {M.~C.}\ \bibnamefont {Schaaf}},\ }\href {https://doi.org/10.22323/1.449.0657} {\bibfield  {journal} {\bibinfo  {journal} {PoS}\ }\textbf {\bibinfo {volume} {EPS-HEP2023}},\ \bibinfo {pages} {657} (\bibinfo {year} {2024})},\ \Eprint {https://arxiv.org/abs/2401.12778} {arXiv:2401.12778 [hep-ph]} \BibitemShut {NoStop}%
\bibitem [{\citenamefont {B{\"u}ndgen}\ \emph {et~al.}(2025)\citenamefont {B{\"u}ndgen}, \citenamefont {Harlander}, \citenamefont {Klein},\ and\ \citenamefont {Schaaf}}]{Bundgen:2025utt}%
  \BibitemOpen
  \bibfield  {author} {\bibinfo {author} {\bibfnamefont {L.}~\bibnamefont {B{\"u}ndgen}}, \bibinfo {author} {\bibfnamefont {R.~V.}\ \bibnamefont {Harlander}}, \bibinfo {author} {\bibfnamefont {S.~Y.}\ \bibnamefont {Klein}},\ and\ \bibinfo {author} {\bibfnamefont {M.~C.}\ \bibnamefont {Schaaf}},\ }\href {https://doi.org/10.1016/j.cpc.2025.109662} {\bibfield  {journal} {\bibinfo  {journal} {Comput. Phys. Commun.}\ }\textbf {\bibinfo {volume} {314}},\ \bibinfo {pages} {109662} (\bibinfo {year} {2025})},\ \Eprint {https://arxiv.org/abs/2501.04651} {arXiv:2501.04651 [hep-ph]} \BibitemShut {NoStop}%
\bibitem [{\citenamefont {Vladimirov}(1980)}]{Vladimirov:1979zm}%
  \BibitemOpen
  \bibfield  {author} {\bibinfo {author} {\bibfnamefont {A.~A.}\ \bibnamefont {Vladimirov}},\ }\href {https://doi.org/10.1007/BF01018394} {\bibfield  {journal} {\bibinfo  {journal} {Theor. Math. Phys.}\ }\textbf {\bibinfo {volume} {43}},\ \bibinfo {pages} {417} (\bibinfo {year} {1980})}\BibitemShut {NoStop}%
\bibitem [{\citenamefont {Misiak}\ and\ \citenamefont {Munz}(1995)}]{Misiak:1994zw}%
  \BibitemOpen
  \bibfield  {author} {\bibinfo {author} {\bibfnamefont {M.}~\bibnamefont {Misiak}}\ and\ \bibinfo {author} {\bibfnamefont {M.}~\bibnamefont {Munz}},\ }\href {https://doi.org/10.1016/0370-2693(94)01553-O} {\bibfield  {journal} {\bibinfo  {journal} {Phys. Lett. B}\ }\textbf {\bibinfo {volume} {344}},\ \bibinfo {pages} {308} (\bibinfo {year} {1995})},\ \Eprint {https://arxiv.org/abs/hep-ph/9409454} {arXiv:hep-ph/9409454} \BibitemShut {NoStop}%
\bibitem [{\citenamefont {Chetyrkin}\ \emph {et~al.}(1998{\natexlab{b}})\citenamefont {Chetyrkin}, \citenamefont {Misiak},\ and\ \citenamefont {Munz}}]{Chetyrkin:1997fm}%
  \BibitemOpen
  \bibfield  {author} {\bibinfo {author} {\bibfnamefont {K.~G.}\ \bibnamefont {Chetyrkin}}, \bibinfo {author} {\bibfnamefont {M.}~\bibnamefont {Misiak}},\ and\ \bibinfo {author} {\bibfnamefont {M.}~\bibnamefont {Munz}},\ }\href {https://doi.org/10.1016/S0550-3213(98)00122-9} {\bibfield  {journal} {\bibinfo  {journal} {Nucl. Phys. B}\ }\textbf {\bibinfo {volume} {518}},\ \bibinfo {pages} {473} (\bibinfo {year} {1998}{\natexlab{b}})},\ \Eprint {https://arxiv.org/abs/hep-ph/9711266} {arXiv:hep-ph/9711266} \BibitemShut {NoStop}%
\bibitem [{\citenamefont {Degrassi}\ \emph {et~al.}(2012)\citenamefont {Degrassi}, \citenamefont {Di~Vita}, \citenamefont {Elias-Miro}, \citenamefont {Espinosa}, \citenamefont {Giudice}, \citenamefont {Isidori},\ and\ \citenamefont {Strumia}}]{Degrassi:2012ry}%
  \BibitemOpen
  \bibfield  {author} {\bibinfo {author} {\bibfnamefont {G.}~\bibnamefont {Degrassi}}, \bibinfo {author} {\bibfnamefont {S.}~\bibnamefont {Di~Vita}}, \bibinfo {author} {\bibfnamefont {J.}~\bibnamefont {Elias-Miro}}, \bibinfo {author} {\bibfnamefont {J.~R.}\ \bibnamefont {Espinosa}}, \bibinfo {author} {\bibfnamefont {G.~F.}\ \bibnamefont {Giudice}}, \bibinfo {author} {\bibfnamefont {G.}~\bibnamefont {Isidori}},\ and\ \bibinfo {author} {\bibfnamefont {A.}~\bibnamefont {Strumia}},\ }\href {https://doi.org/10.1007/JHEP08(2012)098} {\bibfield  {journal} {\bibinfo  {journal} {JHEP}\ }\textbf {\bibinfo {volume} {08}},\ \bibinfo {pages} {098}},\ \Eprint {https://arxiv.org/abs/1205.6497} {arXiv:1205.6497 [hep-ph]} \BibitemShut {NoStop}%
\bibitem [{\citenamefont {Kurkela}\ \emph {et~al.}(2010)\citenamefont {Kurkela}, \citenamefont {Romatschke},\ and\ \citenamefont {Vuorinen}}]{Kurkela:2009gj}%
  \BibitemOpen
  \bibfield  {author} {\bibinfo {author} {\bibfnamefont {A.}~\bibnamefont {Kurkela}}, \bibinfo {author} {\bibfnamefont {P.}~\bibnamefont {Romatschke}},\ and\ \bibinfo {author} {\bibfnamefont {A.}~\bibnamefont {Vuorinen}},\ }\href {https://doi.org/10.1103/PhysRevD.81.105021} {\bibfield  {journal} {\bibinfo  {journal} {Phys. Rev. D}\ }\textbf {\bibinfo {volume} {81}},\ \bibinfo {pages} {105021} (\bibinfo {year} {2010})},\ \Eprint {https://arxiv.org/abs/0912.1856} {arXiv:0912.1856 [hep-ph]} \BibitemShut {NoStop}%
\bibitem [{\citenamefont {Thorne}(2006)}]{Thorne:2006qt}%
  \BibitemOpen
  \bibfield  {author} {\bibinfo {author} {\bibfnamefont {R.~S.}\ \bibnamefont {Thorne}},\ }\href {https://doi.org/10.1103/PhysRevD.73.054019} {\bibfield  {journal} {\bibinfo  {journal} {Phys. Rev. D}\ }\textbf {\bibinfo {volume} {73}},\ \bibinfo {pages} {054019} (\bibinfo {year} {2006})},\ \Eprint {https://arxiv.org/abs/hep-ph/0601245} {arXiv:hep-ph/0601245} \BibitemShut {NoStop}%
\end{thebibliography}%

\end{document}